# Transforming Blood Cell Detection and Classification with Advanced Deep Learning Models: A Comparative Study

Shilpa Choudhary[1], Sandeep Kumar[2], Pammi Sri Siddhaarth[3], Guntu Charitasri[4], Monali Gulhane[5], Nitin Rakesh[6], Feslin Anish Mon[7], Amal Al-Rasheed[8], Masresha Getahun[9], Ben Othman Soufiene[10]

Email_id:
[1]shilpachoudhary2020@gmail.com, [2]er.sandeepsahratia@gmail.com [3]srisiddhaarthp@gmail.com, [4]charitasrig@gmail.com, [5]monali.gulhane4@gmail.com, [6]nitin.rakesh@gmail.com

Corresponding Authors: masreshaggetahun@gmail.com

## Abstract

The detection and classification of blood cells are important in diagnosing and monitoring a variety of blood-related illnesses, such as anemia, leukemia, and infection, all of which may cause significant mortality. Accurate blood cell identification has a high clinical relevance in these patients because this would help to prevent false-negative diagnosis and to treat them in a timely and effective manner, thus reducing their clinical impacts.Our research aims to automate the process and eliminate manual efforts in blood cell counting. While our primary focus is on detection and classification, the outputs generated by our approach can be useful for disease prediction. This follows a two-step approach, where YOLO-based detection is first performed to locate blood cells, followed by classification using a hybrid CNN model to ensure accurate identification.

We conducted a thorough and extensive comparison with other state-of-the-art models, including MobileNetV2, ShuffleNetV2, and DarkNet, for blood cell detection and classification. In terms of real-time performance, YOLOv10 outperforms other object detection models with better detection rates and classification accuracy. But MobileNetV2 and ShuffleNetV2 are more computationally efficient, which becomes more appropriate for resource-constrained environments. In contrast, DarkNet outperformed in terms of feature extraction performance, and the fine blood cell type classification.

Additionally, an annotated blood cell data set was generated for this study. A diverse set of blood cell images with fine-grained annotations is contained in this dataset to make it useful for deep learning models training and evaluation. Because the present dataset will be an important resource for researchers and developers working on automatic blood cell detection and classification systems, we will make it publicly available under the open-access nature in order to accelerate the collaboration and progress in this field.

## 1. Introduction

Detecting and classifying blood cells is an important step for diagnosing a range of hematological disorders, such as anemia, leukemia, infections and other disorders associated with the blood. Traditionally, this involved examination under a microscope by pathologists, a process that can be human-labor intensive, subjective, and error-prone. Moreover, the increasing range of diagnostic tests has escalated a growing demand for automated, high throughput solutions that are cost-effective, require a rapid turnaround time and provide reliable results. This new model built on recent developments in deep learning, namely convolutional neural networks (CNNs), and radically changed the landscape around medical imaging and diagnostics [1-2].

In the context of blood cell analysis, CNNs are capable of identifying and classifying blood cell morphologies (such as red blood cells, white blood cells, or platelets) with little or no human involvement. Such automation minimizes human intervention, increases the accuracy of diagnostics, and fast tracks the process. Deep learning shows promise for automating blood cell analysis, but there are still challenges to address. The most prevalent drawbacks of existing techniques are the lacking aspects of real-time, robustness and generalizability. In addition, absence of extensive comparison studies on performance of recent deep learning architectures for detection and classification of blood cells impedes selection of the appropriate models for clinical application. This is compounded by restricted access to large quality annotated datasets which stifle innovation and descent into ineffectual progress. The use of advanced architectures (e.g., ResNet, EfficientNet, and transformers), transfer learning, data augmentation, and ensemble methods have made these models much more robust and generalizable. Such methods facilitate the detection of subtle morphological variations in blood cell structures, leading to clinical implementations [4-9].

Despite this, the lack of availability of some datasets, class imbalance and the difficulty computing the necessary high-dimensional covers or mines themselves, and incorporating interpretability or generalizability across differing imaging systems, have thwarted the widespread success of many of these approaches and models. Solving these issues is pivotal in order to harness the maximal powers of these deep learning models on blood cell detection and classification [10-13].

Recent research has demonstrated the effectiveness of deep learning in medical image analysis, particularly for blood cell detection and classification. Optimized CNN models have been used for cancer diagnosis, highlighting techniques that enhance classification accuracy in real-time detection tasks. Knowledge distillation approaches have been applied to leukemia detection, aligning with YOLOv10's efficiency in handling medical datasets. Additionally, ensemble models with meta-heuristic algorithms have been utilized for tumor classification, emphasizing the benefits of hybrid architectures like MobileNetV2, ShuffleNetV2, and DarkNet for precision medicine applications. These studies reinforce the potential of deep learning models for accurate and efficient blood cell classification [14-18].

Advancements in deep learning have significantly improved medical image analysis across various domains, including cancer detection and brain stroke segmentation. Hybrid CNN-ViT models have demonstrated effectiveness in early disease diagnosis, aligning with the precision and efficiency required for real-time blood cell classification. Attention-enhanced architectures have further refined feature extraction for complex medical imaging tasks, similar to how YOLOv10 and lightweight models like MobileNetV2 and ShuffleNetV2 optimize detection. Additionally, U-Net-based segmentation methods emphasize the importance of accurate region localization, which is crucial for blood cell detection and classification in precision medicine applications. [19-24].

The automated detection and classification of blood cells not only accelerate the diagnostic process while increasing the accuracy of diagnosis at an early stage, but also serve as the basis for the pathophysiological study of diseases, like anemia, leukemia, and infections. With the support of deep learning and computer vision, the health systems can move towards an automatic, precise and scalable diagnostic workflow, which was performed with the traditional manual inspection methods.

We have compared the traditional diagnostic methods and high-order deep neural network (DNN) architectures. While traditional methods rely on human intelligence, resulting in less accurate, time-consuming, and variable processes, DNNs provide faster processing and better results. Although DNNs require significant investment in business and technical analysis, they offer better adaptation to new data, scalability, and lower costs over time. Overall, DNNs represent a significant advance in diagnostics, providing many improvements over traditional methods. The comparison between the traditional method and DNN is summarized in Table 1.

**Table 1:** Comparative Analysis between Traditional Methods and Advanced DNN Architectures

| Aspect | Traditional Methods | Advanced DNN Architectures |
|---|---|---|
| Accuracy | Moderate, highly dependent on the haematologist's expertise | High, improves with more epochs and model training |
| Time Efficiency | Time-consuming due to manual analysis | Faster due to automated processing and real-time capabilities |
| Error Rate | Prone to human error | Lower error rate with well-trained models |
| Consistency | Variable, dependent on haematologist's experience | High, consistent results across different datasets |
| Adaptability | Limited flexibility to new patterns or data | High adaptability through retraining and updating |
| Resource Requirement | Requires skilled radiologists | Requires computational resources and data scientists |
| Scalability | Limited scalability due to reliance on human resources | Highly scalable with appropriate infrastructure |
| Clinical Validation | Well-established with long-term usage | Requires extensive validation and regulatory approval |
| Cost | High due to the need for specialized personnel | Initial setup costs can be high but potentially lower in the long term due to automation. |

One of the most advanced real-time object detection systems is YOLO, which is known for its high speed and accuracy. Among the variants released in it exists YOLOv10, which improves appreciably over its predecessors. This paper presents the application of the YOLOv10 framework for detecting and classifying blood cells from microscopic images. We used a robust Roboflow dataset containing annotated images with many blood cell types. The YOLOv10 model we trained at epochs 10, 50, and 100, with batch size 16 and image resolution of 640 × 640 pixels, had to balance computational efficiency against detection accuracy.

The aim of this study is to cope with these challenges by presenting a novel real-time system for blood cell detection and classification using YOLOv10 that outperforms the current best models (MobileNetV2, ShuffleNetV2, DarkNet) and lays down benchmark performance against which automated segmentation and classification of blood cells can be compared against, while introducing a novel dataset to fast-track future research in automated haematological analysis.

In this work we make the following main contributions:

- Building an End-to-end Real-time Detection and Classification System.
- Performance Comparison of State-of-the-Art Models.
- Created and formalized a highly specific dataset for blood cell research, thus being a scientific data deep-dive content for researchers.

The rest of the document is structured as follows: The remainder of the paper is structured as follows: Section 2 provides an overview of the relevant work, while Section 3 outlines the methodology that has been presented. Sections 4 &5 contain the findings and conclusion.

## 2. Literature Review

The research paper focuses on blood cell detection and classification, which is critical in diagnosing and treating many haematological disorders like anaemia, leukaemia, and infections. Traditionally, methods of blood cell analysis require much involvement by manual inspection from experts, usually haematologists or laboratory technicians. These haematologists apply microscopic examination techniques in classifying various types of blood cells based on their morphology. Though the method has been a standard for years, it is time-consuming, subjective, and greatly dependent upon the expertise of the person doing the analysis.

The paper [1] proposes a study for early and accurate detection of blood cancer, or haematological malignancy, which is essential for proper treatment and better patient outcomes. Deep learning has recently undergone many advances in providing powerful tools to enhance blood cancer detection with various architectures, datasets, and preprocessing techniques. The research [2] investigates the application of Convolutional Neural Networks in blood cell classification and proves that with their superior accuracy and efficiency, they outperform traditional manual methods. CNNs do the automation of extensive datasets processing, hence improving diagnostic speed and reliability in clinical settings. The study [3] mentioned a new architecture of CNN-based methods for detecting and counting blood cells, where VGG-16 acts as the backbone, with feature fusion and the mechanism of block attention, namely, CBAM. The highest Recall was obtained for RBC detection with Model 3, which measured 82.3 and 86.7% under confidence scores of 0.9 and 0.8, respectively. Model 1 worked well for WBC detection and achieved 76.1% precision with a recall of 95% at a confidence threshold of 0.9 and 69.1% with a recall of 96.4% at 0.8. Platelet detection could have been better, especially for platelets that were in clusters. The paper [4] presents modern communication techniques to identify diseases earlier and overcome the drawbacks of the manual technique. It classifies RBCs in the microscopic image using an ECNN model with 95% accuracy and 0.93 precision. Better performance is shown with the existing standard methods, and future works are to touch other deeper learning architectures like RNN and LSTM for further accuracy and precision. The research [5] presents a new three-stage pipeline where an SNN combined with a CNN detects malaria in blood smears to within a global accuracy of 93.72% in RBC segmentation. While the CNN was relatively accurate during validation about classification between the infected and non-infected RBCs, it did very poorly during generalization on the test set, bringing out areas for future improvements. This paper [6] proposes a hybrid classification approach for White Blood Cell Leukemia by integrating VGGNet for feature extraction with a statistically enhanced Salp Swarm Algorithm as a feature optimization technique. This provides better accuracy and offers a remarkable reduction in features from 25,000 to 1,000, reducing the complexity and improving the efficiency of the existing models for the same dataset. The paper [7] proposes three deep learning models for classifying erythrocytes under the categories of circular and elongated, which means normal and sickle cell, respectively, with other blood content using traditional and parallel architectures of CNNs. Such models demonstrate prominent performance improvement and are more robust to classification accuracy, especially with transfer learning and data augmentation methods, outperforming any current method in sickle cell disease classification and opening the scope for application in white blood disease classification. This paper [8] presents an improved RCD algorithm for automating segmentation and counting red and white blood cells. Some key improvements include 8-neighbor component initialization, enhanced circle selection, and irregular and overlapping cell handling. In the experimental tests, it achieved a high accuracy with a precision rate of 89.7% and a recall rate of 98.4% against expert-determined ground truth.

## 3. Proposed Methodology

In this research, the detection of the blood cells was approached by employing a YOLOv10 model and several convolutional neural network architectures for classification. The detection process involves training a YOLOv10 model for 100 epochs. A detailed Flow chart of the proposed work is shown in Figure 1.

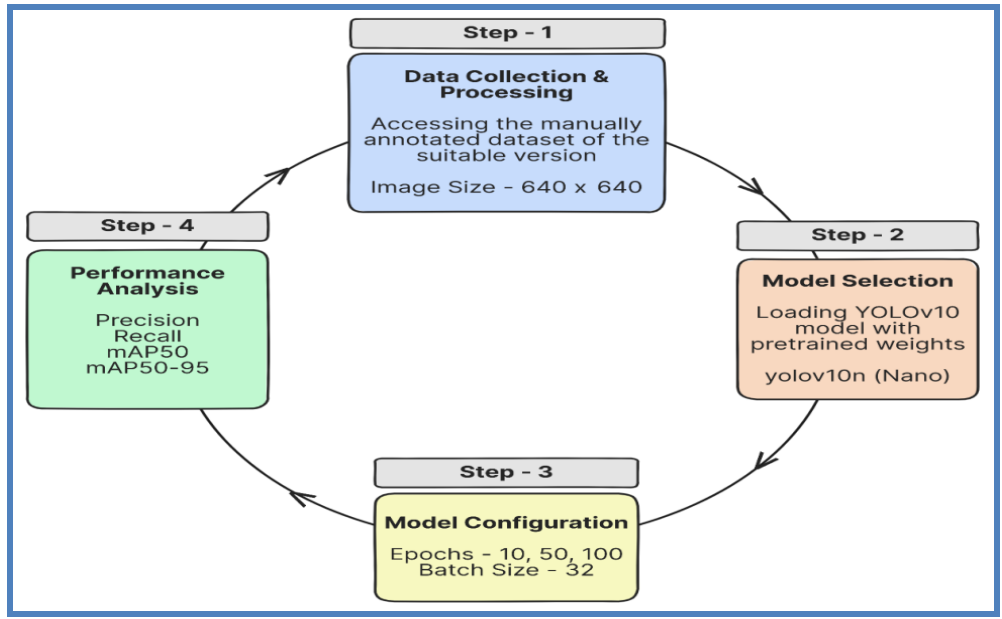

**Figure 1:** Steps for Blood Cell Detection Using YOLOv10

### 3.1 Data Collection

The dataset [9] comprises 17,092 images of individual normal cells captured using the CellaVision DM96 analyzer in the Core Laboratory at the Hospital Clinic of Barcelona. Under sampling was performed on the data to get a uniform and balanced dataset with the class distribution shown in Table 2, and sample images are shown in Figure 2.

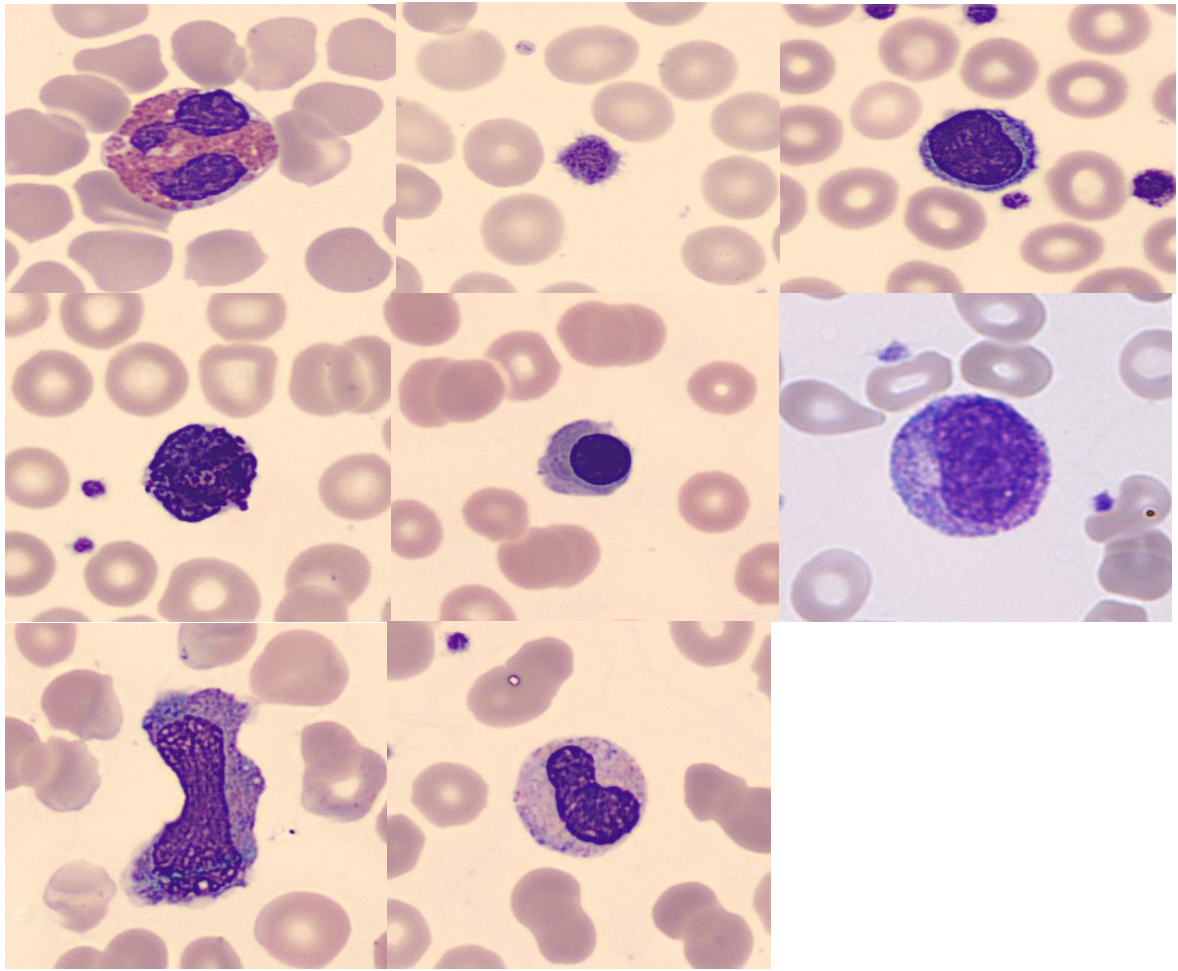

**Figure 2:** Images of Eosinophil, Platelet, Lymphocyte, Basophil, Erythroblast, IG, Monocyte, Neutrophil

**Table 2:** Distribution of Blood Cell Types in Dataset

| **Blood Cell Type** | Data Size |
|---|---|
| Basophils | 1218 |
| Eosinophils | 3117 |
| Erythroblasts | 1551 |
| Immunoglobulins | 2895 |
| Lymphocytes | 1214 |
| Monocytes | 1420 |
| Neutrophils | 3329 |
| Platelets | 2348 |

### 3.1 Data Preprocessing

The dataset annotation through Roboflow has also contributed to the data preprocessing stage. This ensures that the dataset about the blood cells is correctly labelled, providing a very reliable basis for training. The dataset was split into 70% training, 20% testing, and 10% validation to ensure a well-balanced evaluation process. After the annotation, a proper dataset version is selected for further usage. The preprocessing process starts by annotating a raw dataset to create an annotated version **$D_{annot}$=Annotate ($D_{raw}$)**, ensuring that blood cells are accurately labelled. A specific version of this annotated dataset is then chosen for continued use, denoted as **$D_{final}$=V($D_{annot}$)**. This version is then exported as a ZIP file, **$D_{export}$=Export ($D_{final}$, ZIP)**, which contains the images and associated configuration files **C($D_{export}$)=ExtractConfig($D_{export}$)** needed for training the YOLOv10 model. Data augmentation methods are also applied to improve the dataset, resulting in an augmented **Daug = Augment (Dfinal)**, thereby establishing a robust foundation for the model's training and evaluation stages. Transformations such as random horizontal flips and normalization with mean and standard deviation values were applied, and three augmented variants were added to enhance dataset diversity.

### 3.2 Optimizing YOLOv10n for Blood Cell Detection: Training and Performance Analysis

The model was initialized with the pre-trained weights from YOLOv10n, so it could begin at a good initialization point for fine-tuning on a blood cell detection dataset and learning prior knowledge to increase training efficiency, as shown in Figure 3.YOLOv10n was selected over YOLOv10 due to its lightweight architecture and lower parameter count, making it more suitable for transfer learning on the given blood cell dataset. Since YOLOv10n is designed for scenarios with limited data, it provides an efficient solution without the computational overhead of larger models. Pretrained weights lessen convergence time by providing the model with a robust initialization, allowing it to utilize prior knowledge rather than learning from scratch. Given the nature of blood cell detection, a lightweight model like YOLOv10n is sufficient for effective feature extraction and classification, eliminating the need for more complex architectures with higher parameter counts. In contrast, Vision Transformers (ViTs) and other hybrid CNN architectures typically have significantly larger parameter counts, making them computationally demanding and less suitable for lightweight applications such as this. The number of epochs was 100, essential for efficient learning and convergence. The batch size used here was 32, which balances memory usage and speed during the training process, making data processing efficient, as shown in Table 3. Furthermore, the input images were resized to 640x640 pixels, putting the training in a standard process to maintain uniformity throughout the dataset. The training execution was performed on Google Colab, which provided the computational resources needed for the training process.

# Algorithm: Proposed Work

The initial weights for the YOLOv10n model are represented as:

$$W_{init} = W_0 + \Delta W$$

$W_{init}$: Initial model weights after transfer learning, $W_0$: Pre-trained weights from YOLOv10n, $\Delta W$: Weight adjustments during fine-tuning.

The input image I is resized to a uniform size of 640×640pixels:

$$I_{resized} = f_{resized}(I_{original}, (640, 640))$$

$f_{resized}$: Function that resizes the image, $I_{original}$: Original input image, $I_{resized}$: Resized image. Let: E: Number of epochs, B: Batch size, α: Learning rate.

$$E=100, B=32, \alpha=\alpha_{adaptive}$$

For each image $I_{resized}$, the model predicts:

- Bounding boxes $\hat{B} \in R^4$
- Class probabilities $\hat{C} \in R^{N_{class}}$
- Objectness scores $\hat{O} \in R$

The predicted output vector YpredY_{\text{pred}}Ypred is:

$$Y_{pred} = (\hat{B}, \hat{C}, \hat{O})$$

The total loss function L consists of three components:

$$\mathcal{L} = \lambda_{bbox} \cdot \mathcal{L}_{bbox} + \lambda_{class} \cdot \mathcal{L}_{class} + \lambda_{obj} \cdot \mathcal{L}_{obj}$$

$\lambda_{bbox}, \lambda_{class}, \lambda_{obj}$: Weight coefficients for balancing the losses.

The bounding box loss uses Mean Squared Error (MSE):

$$\mathcal{L}_{bbox} = \frac{1}{B} \sum_{i=1}^{B} \left((x_i - \hat{x}_i)^2 + (y_i - \hat{y}_i)^2 + (w_i - \hat{w}_i)^2 + (h_i - \hat{h}_i)^2\right)$$

- $\hat{x}_i, \hat{y}_i, \hat{w}_i, \hat{h}_i$: Predicted bounding box coordinates.

The class probability loss uses Cross-Entropy:

$$\mathcal{L}_{class} = -\frac{1}{B}\sum_{i=1}^{B}\sum_{c=1}^{N_{class}} C_{i,c}\log(\widehat{C_{i,c}})$$

$C_{i,c}$: Ground truth class label (1 if class ccc is present, 0 otherwise).

$\widehat{C_{i,c}}$: Predicted probability for class ccc.

The objectness loss uses Binary Cross-Entropy:

$$\mathcal{L}_{obj} -\frac{1}{B}\sum_{i=1}^{B}\left(O_i \log(\hat{O}_i) + (1 - O_i)\log(1 - \hat{O}_i)\right)$$

- $O_i$: Ground truth objectness score (1 if object is present, 0 otherwise).
- $\hat{O}_i$: Predicted objectness score.

The weights are updated iteratively using the Adam optimization rule:

$$m_t = \beta_1 \cdot m_{t-1} + (1 - \beta_1) \cdot \frac{\partial \mathcal{L}}{\partial W_t}$$

$$v_t = \beta_2 \cdot v_{t-1} + (1 - \beta_2) \cdot \left(\frac{\partial \mathcal{L}}{\partial W_t}\right)^2$$

$$\widehat{m}_t = \frac{m_t}{1 - \beta_1^t}, \quad \hat{v}_t = \frac{v_t}{1 - \beta_2^t}$$

$$W_{t+1} = W_t - \alpha \frac{\widehat{m}_t}{\sqrt{\hat{v}_t} + \epsilon}$$

- $m_t, v_t$: Moving averages of gradients and squared gradients.
- $\beta_1, \beta_2$: Hyperparameters for moving averages.
- $\epsilon$: Small constant for numerical stability.

**Table 3:** Model Training Parameters for YOLOv10

| Epochs | 100 |
|---|---|
| Batch Size | 32 |
| Image Size | 640 x 640 |

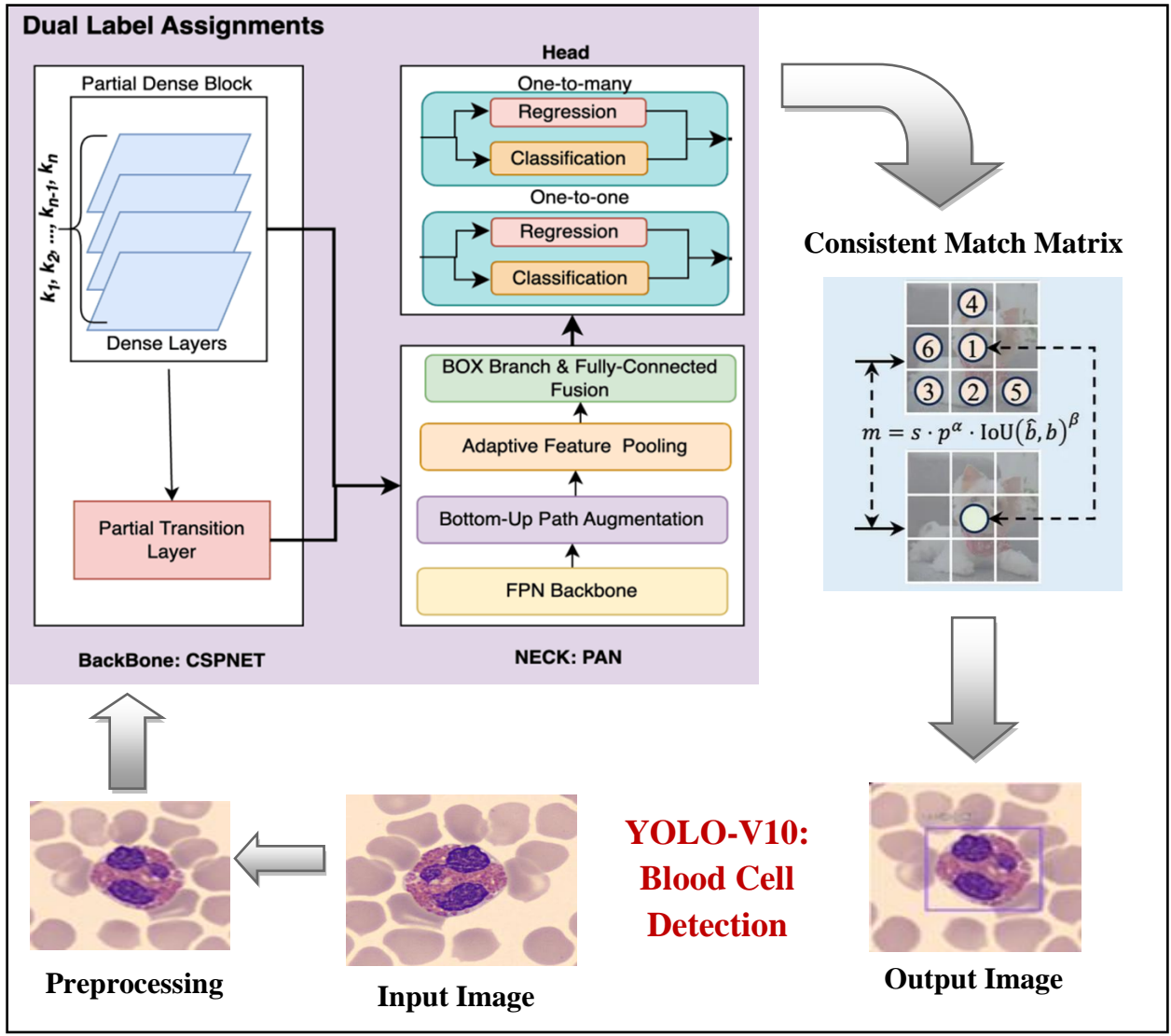

**Figure 3:** Flow chart of blood cell detection

Regarding data loading, augmentation, and model optimization, the Ultralytics YOLOv5 and later versions have built-in support for handling all the tasks, making the whole training workflow much more accessible. During the training process, the primary performance metrics were Precision, Recall, and the F1 score. The performance metrics have contributed to evaluating its model accuracy with blood cell detection, thus ascertaining its performance at every epoch. The best weights, training curves, confusion matrices, and other relevant metrics that resulted from training were all saved for later analysis. This way, a detailed assessment is obtained, which further helps fine-tune and improve the model. TensorBoard was employed, which provided all-in-one visualization tools for different performance metrics and training progress; using it, the YOLO model was evaluated. The detailed description of proposed work is mentioned in the Algorithm 1.

## 3.3 Performance Comparison of CNN Architectures: MobileNetV2, ShuffleNetV2, and Darknet across Epochs and Metrics

In this step, deep architectures were compared using some top-rated models: MobileNetV2, ShuffleNetV2, and Darknet, as shown in Figure 4.

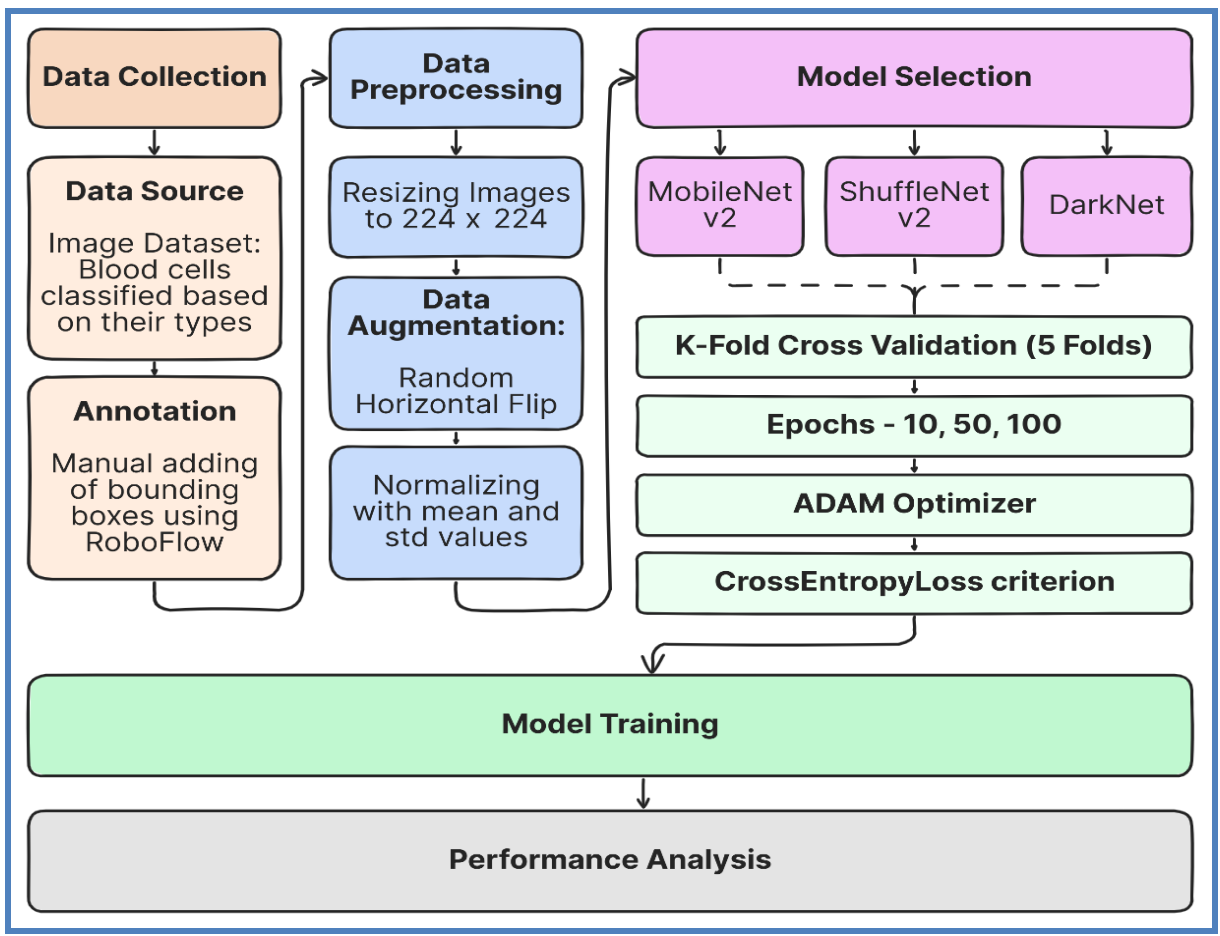

**Figure 4:** Steps for Blood Cell Classification Using Convolutional Neural Network Architectures

These architectures were assessed on a classification task for their performance and suitability. The dataset was partitioned into five folds for cross-validation to ensure the models were tested as comprehensively as possible. This would be important in assessing each model's generalization by training on different subsets of data and validating others. The detailed training process across different epoch counts was done to establish how the number of training epochs affects the performance of the model. In this case, every model was trained for 10, 50, and 100 epochs to see how longer training would affect accuracy and other performance metrics. A constant batch size of 32 was used for training and testing. Adam optimizer was chosen due to its ability to handle sparse gradients, and its adaptive learning rate would be helpful in very complex models. Cross-entropy loss was used to calculate the loss since this is appropriate for tasks involving multiclass classification. During training, several performance metrics, accuracies, precision, recall, and F1 score, were calculated at the end of every epoch. These metrics have been derived using functions from the sci-kit-learn library, including robust tools to evaluate a classification model. These weights were stored after each model training so the saved weights could be used later to benchmark model performance against the test set or even further train any models. The specification of model training parameters is shown in Table 4. After hyper tunning of various parameters outcome of the proposed work for all steps for blood cell detection and classification is shown in the Figure 5.

**Table 4:** Model Training Parameters for CNN Architectures

| | |
|---|---|
| Image Size / Resolution | 224 x 224 |
| Data Augmentation | Random Horizontal Flip and Normalization |
| K-Fold Cross Validation | 5 Folds |
| Epochs | 10, 50, 100 |
| Optimizer | ADAM |
| Loss Function | Cross Entropy |

## 4. Results & Discussions

### 4.1 Experimental Environment

The training was conducted on Google Colab, utilizing an NVIDIA Tesla T4 GPU, which provided the necessary computational resources. The Tesla T4 is equipped with 16GB of GDDR6 memory and optimized for deep learning applications, offering a balance between computational efficiency and performance.Based on the observed training speed during the first epoch, where 354 iterations were processed in approximately 128 seconds, the training duration for 10 epochs took around 21 minutes, while 50 epochs took approximately 1 hour and 46 minutes, and 100 epochs took about 3 hours and 33 minutes. These durations assume a consistent iteration speed and stable GPU performance. The recorded GPU memory usage of 3.02GB indicates efficient memory utilization, ensuring smooth training without exceeding the hardware's memory constraints.TheUltralytics YOLOv10 is used to load data, augment it, and optimize the model. Precision, Recall, and the F1 score were other performance metrics considered in this work. A Tensor board was used to visualize the training progress and performance metrics and undertake a complete evaluation of the YOLO model. For classification, a variety of augmentation techniques were applied during training for the betterment of the model. These include random flips and crops that made the training dataset rich and prevented a model from over-fitting.

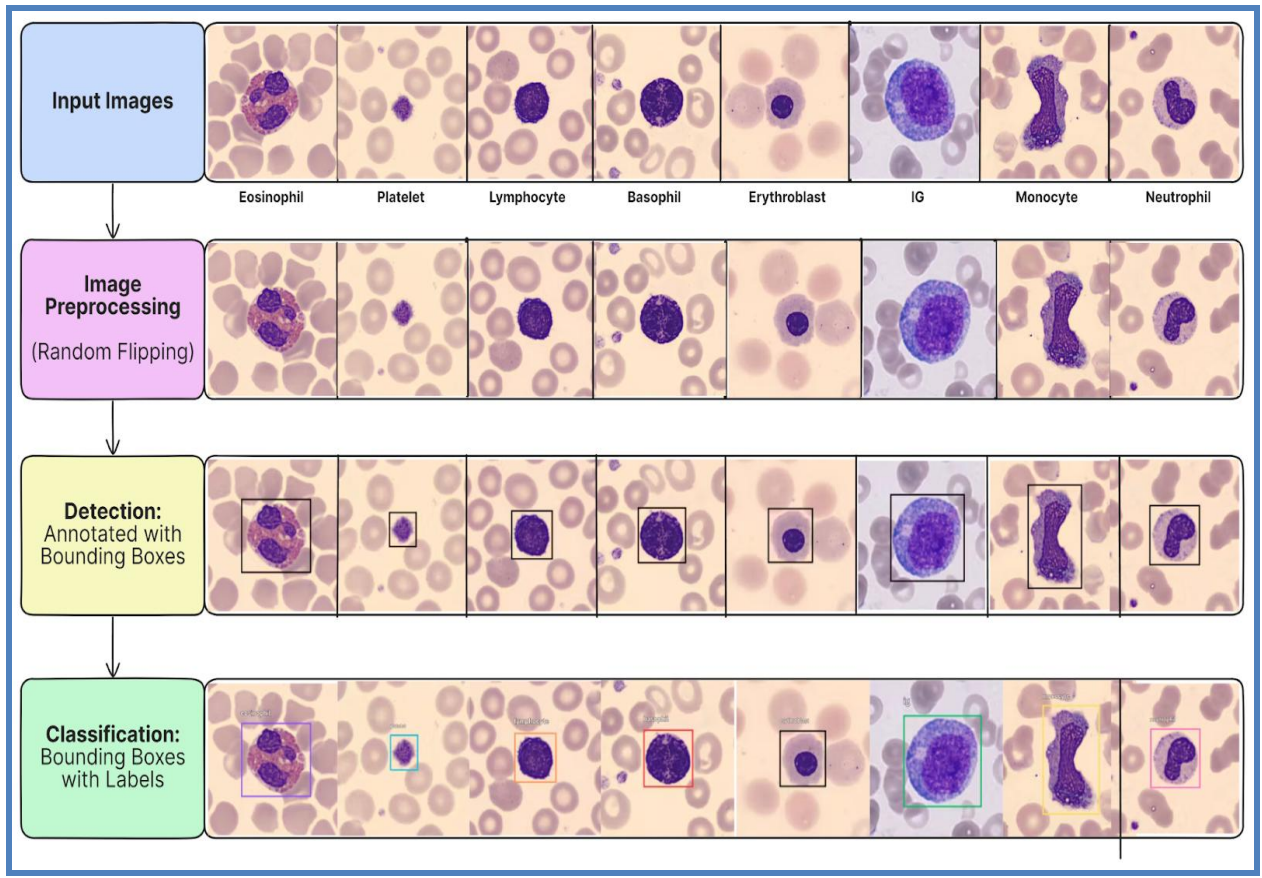

**Figure 5:** Proposed Work Outcomes for all Steps for Blood Cell Detection and Classification

## 4.2 Comparative Analysis of Different Epoch Configurations on YOLOv10 for Blood Cell Detection

   a. Comparative Analysis of epoch configurations on YOLOv10 for Blood Cell Detection for ten epochs

For 10 epochs, the model showed steady progress in blood cell classification. The training losses were 1.51957 for box loss, 0.84895 for class loss, and 2.13515 for DFL loss. As shown in the Table 5, the precision and recall for class B were 0.94895 and 0.93841, respectively. The mAP50 for class B was 0.97634, with an mAP50-95 of 0.78037. The validation losses for box, class, and DFL were 1.61982, 0.70064, and 2.15377, respectively. The learning rates for all parameter groups were 9.0797e-05, indicating the model was in the early stages of training and adjusting its learning rate for improvement. The results are shown in the Figure 6 - Figure 8.

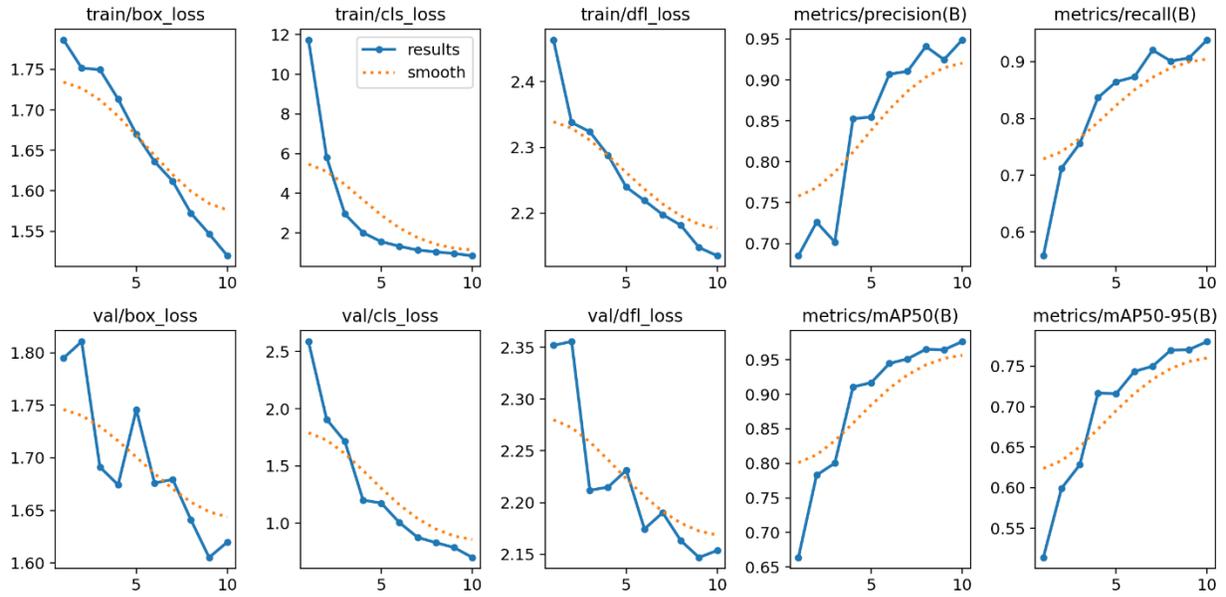

**Figure 6:** The performance of a YOLOv10 model across different metrics over ten epochs.

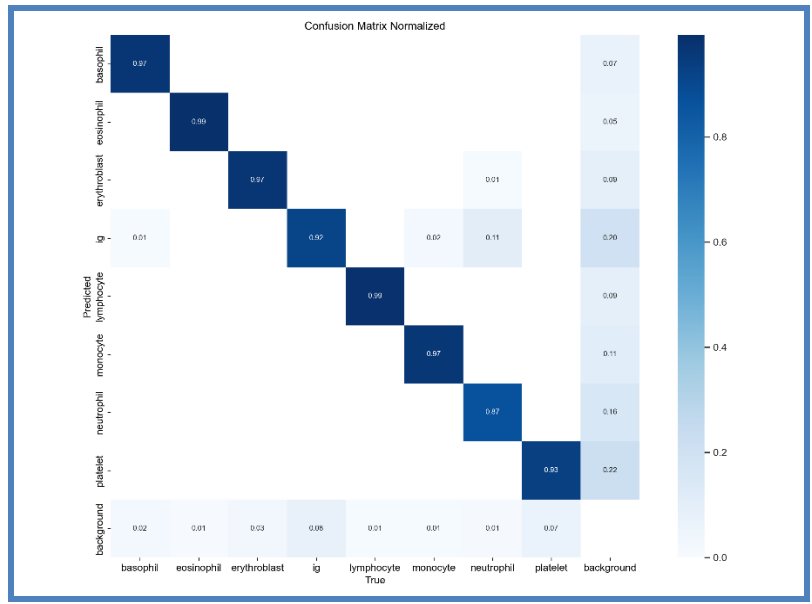

**Figure 7:** The Normalized Confusion Matrix for Multiclass Classification Model (10 epochs).

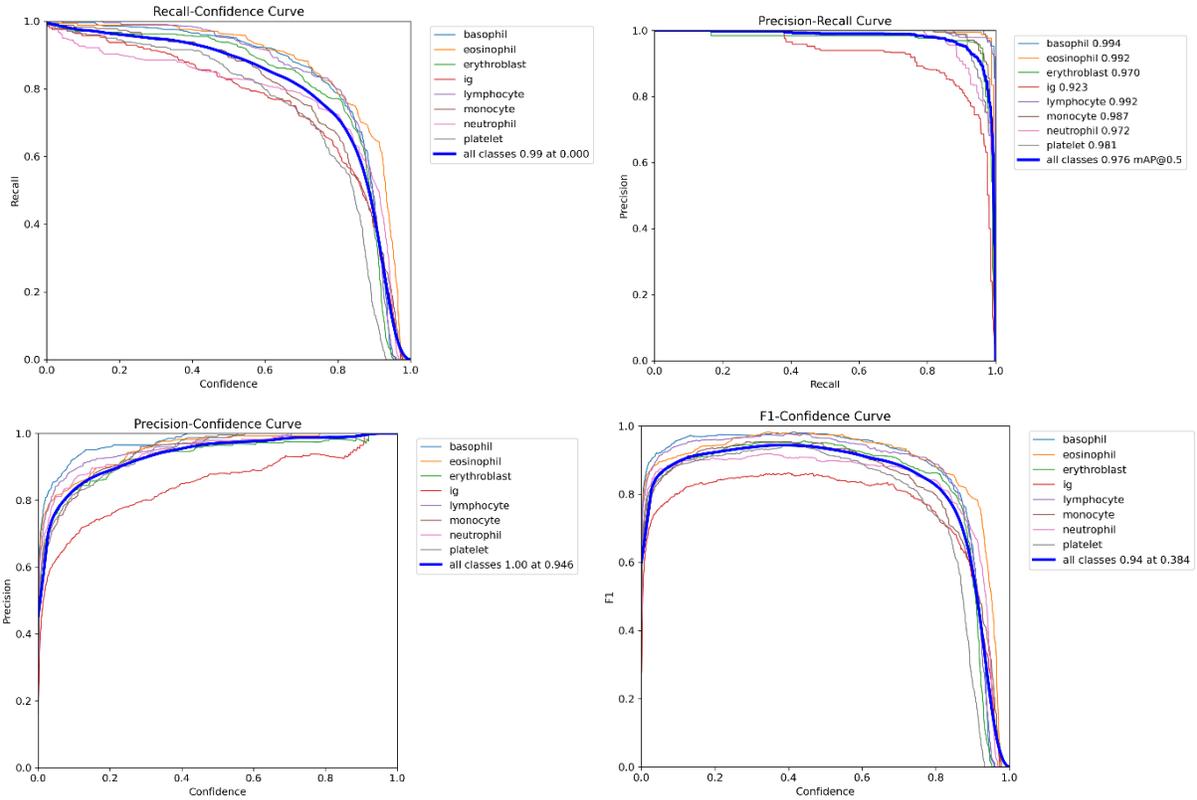

**Figure 8:** The performance evaluation: Recall-Confidence, Precision-Recall, Precision-Confidence and F1-Confidence Curve for Multiclass Classification (10 epochs).

**Table 5:** Performance metrics of the YOLOv10 model for blood cell detection and classification across different cell types. P: Precision, R: Recall, mAP50: mean Average Precision at 50% IoU, mAP 50-95: mean Average Precision across IoU thresholds from 50% to 95% for ten epochs

| Class | Images | Instances | Precision | Recall | mAP50 | mAP50-95 |
|---|---|---|---|---|---|---|
| **all** | 1616 | 1685 | 0.948 | 0.938 | 0.976 | 0.781 |
| **basophil** | 201 | 201 | 0.985 | 0.971 | 0.994 | 0.819 |
| **eosinophil** | 167 | 168 | 0.979 | 0.976 | 0.992 | 0.877 |
| **erythroblast** | 223 | 241 | 0.947 | 0.958 | 0.97 | 0.695 |
| **ig** | 193 | 208 | 0.836 | 0.882 | 0.923 | 0.772 |
| **lymphocyte** | 193 | 196 | 0.96 | 0.986 | 0.992 | 0.785 |
| **monocyte** | 179 | 179 | 0.969 | 0.939 | 0.987 | 0.792 |
| **neutrophil** | 213 | 218 | 0.955 | 0.877 | 0.972 | 0.82 |
| **platelet** | 247 | 274 | 0.957 | 0.916 | 0.981 | 0.684 |

**b. Comparative Analysis of epoch configurations on YOLOv10 for Blood Cell Detection for 50 epochs**

For 50 epochs, the model achieved impressive results in blood cell classification. The training losses were 1.39947 for box loss, 0.4876 for class loss, and 2.13827 for DFL loss. As shown in the Table 6, Figure 9- Figure 11, the model demonstrated high precision and recall for class B, with values of 0.97304 and 0.95988, respectively. The mAP50 for class B was 0.98893, and the mAP50-95 was 0.80721. The validation losses for box, class, and DFL were 1.55685, 0.55655, and 2.23229, respectively. The learning rates for the parameter groups were extremely low, around 2.48234e-05, indicating the model was nearing convergence.

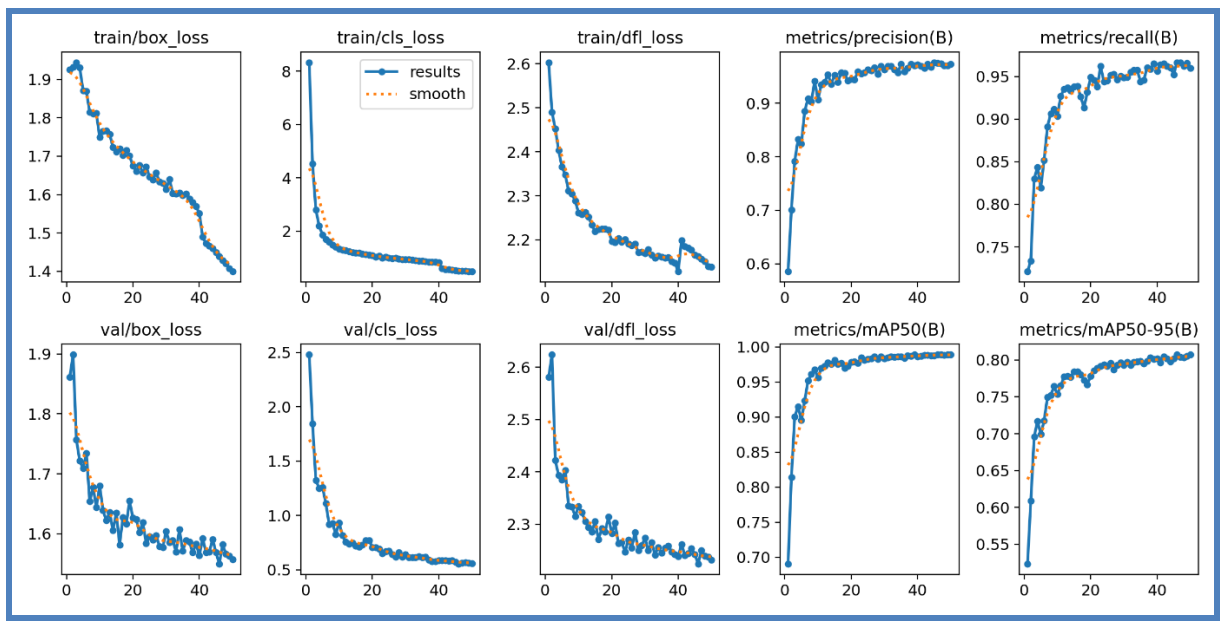

**Figure 9:** The performance of a YOLOv10 model across different metrics over 50 epochs.

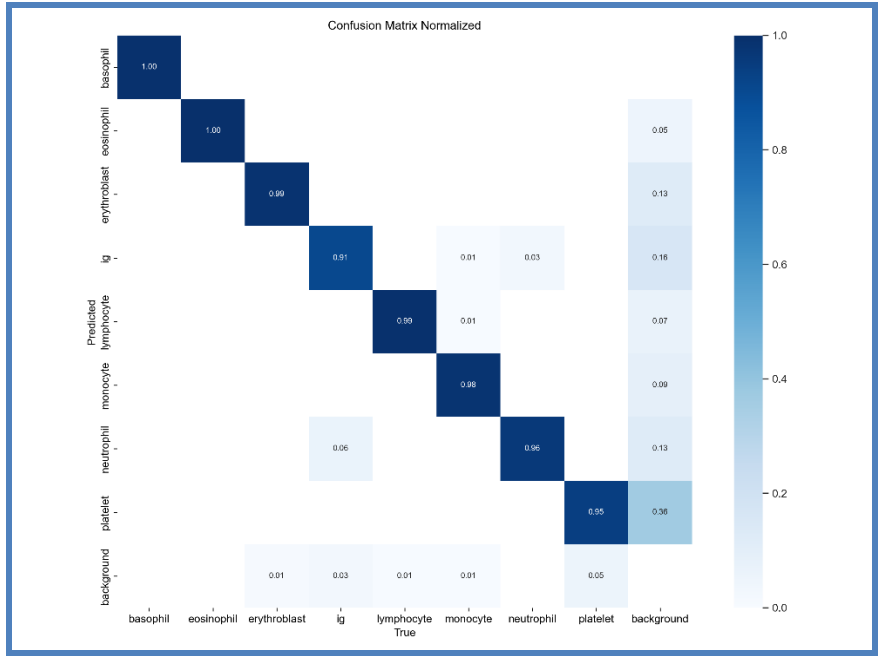

**Figure 10:** The Normalized Confusion Matrix for Multiclass Classification Model (50 epochs).

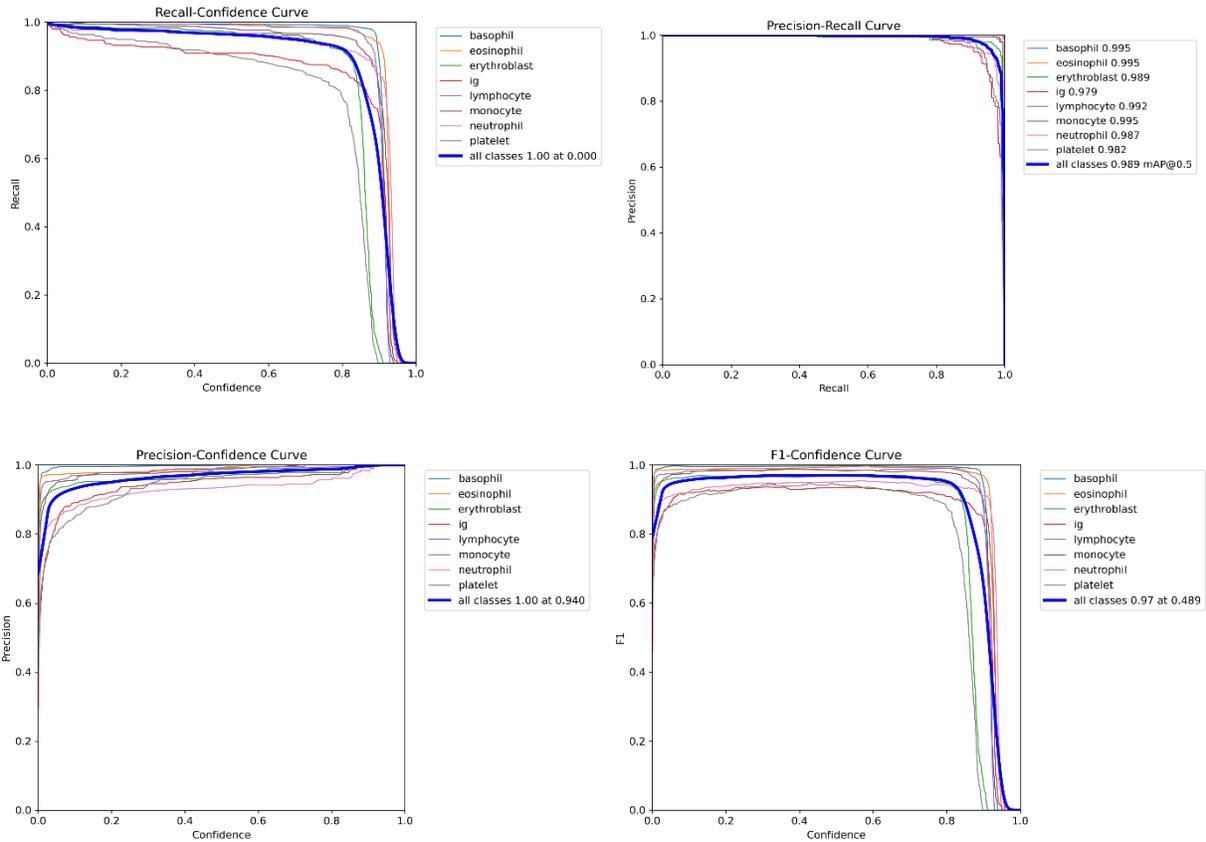

**Figure 11:** The performance evaluation: Recall-Confidence, Precision-Recall, Precision-Confidence and F1-Confidence Curve for Multiclass Classification. (50 epochs).

**Table 6:** Performance metrics of the YOLOv10n model for blood cell detection and classification across different cell types. P: Precision, R: Recall, mAP50: mean Average Precision at 50% IoU, mAP50-95: mean Average Precision across IoU thresholds from 50% to 95% for 50 epochs.

| Class | Images | Instances | Precision | Recall | mAP50 | mAP50-95 |
|---|---|---|---|---|---|---|
| All | 1616 | 1685 | 0.978 | 0.962 | 0.989 | 0.808 |
| basophil | 201 | 201 | 1 | 0.995 | 0.995 | 0.825 |
| eosinophil | 167 | 168 | 0.995 | 0.994 | 0.995 | 0.909 |
| erythroblast | 223 | 241 | 0.971 | 0.967 | 0.989 | 0.747 |
| Ig | 193 | 208 | 0.962 | 0.909 | 0.979 | 0.831 |
| lymphocyte | 193 | 196 | 0.98 | 0.992 | 0.992 | 0.795 |
| monocyte | 179 | 179 | 0.989 | 0.979 | 0.995 | 0.809 |
| neutrophil | 213 | 218 | 0.936 | 0.968 | 0.987 | 0.848 |
| platelet | 247 | 274 | 0.992 | 0.897 | 0.982 | 0.696 |

c. **Comparative Analysis of epoch configurations on YOLOv10 for Blood Cell Detection for 100 epochs**

For 100 epochs, the model continued to show strong performance in blood cell classification. The training losses were 1.3287 for box loss, 0.4344 for class loss, and 2.09103 for DFL loss. As shown in the Figure 12- Figure 14 & Table 7, the precision and recall for class B were high at 0.9827 and 0.9637, respectively. The mAP50 for class B was 0.98913, with an mAP50-95 of 0.80914. The validation losses for box, class, and DFL were 1.56829, 0.54816, and 2.24784, respectively. The learning rates for all parameter groups were very low at 1.65767e-05, indicating that the model was stabilizing and nearing optimal performance.

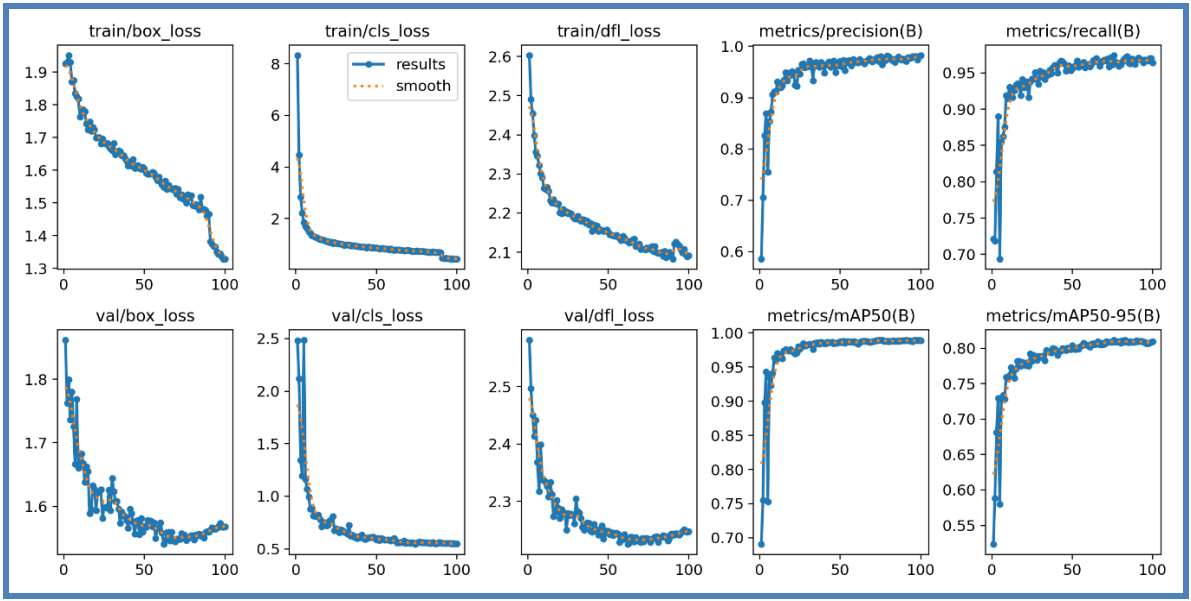

**Figure 12:** The performance of a YOLOv10 model across different metrics over 100 epochs.

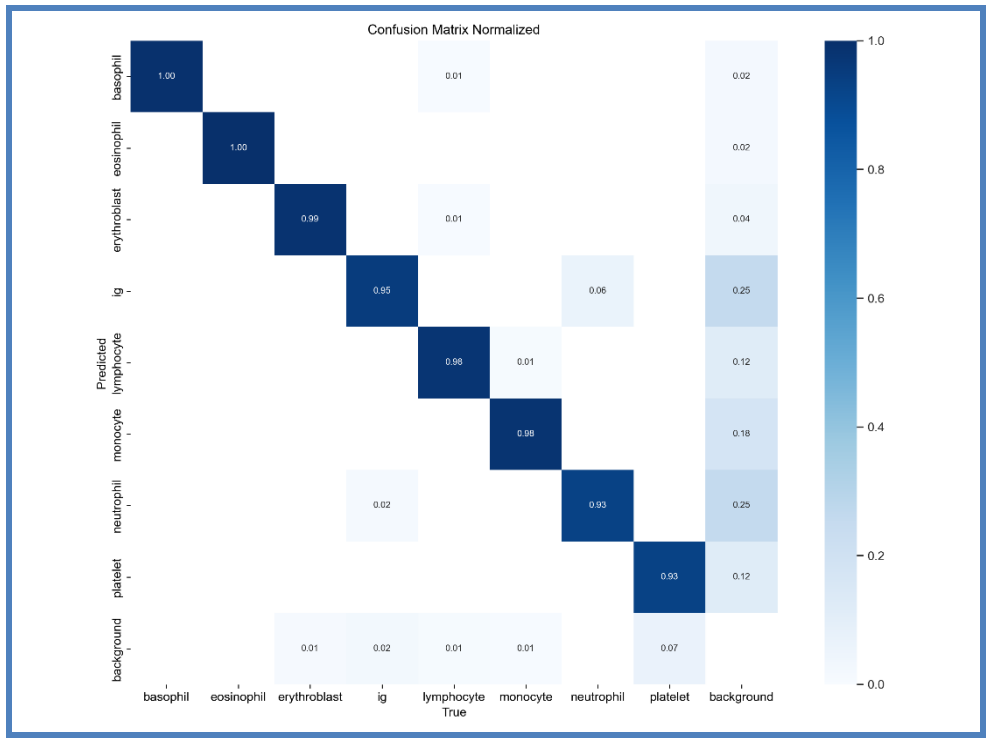

**Figure 13:** The Normalized Confusion Matrix for Multiclass Classification Model (100 epochs).

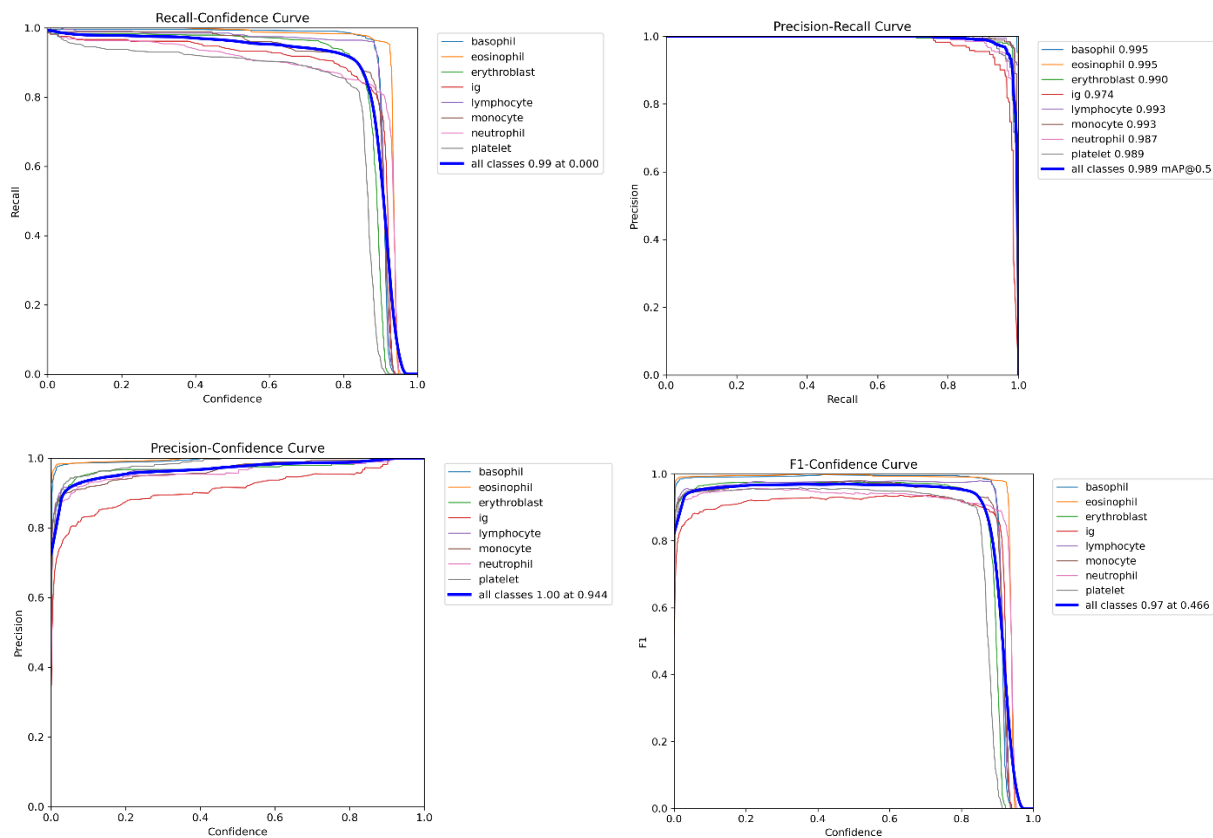

**Figure 14:** The performance evaluation: Recall-Confidence, Precision-Recall, Precision-Confidence and F1-Confidence Curve for Multiclass Classification. (100 epochs).

**Table 7:** Performance metrics of the YOLOv10n model for blood cell detection and classification across different cell types. P: Precision, R: Recall, mAP50: mean Average Precision at 50% IoU, mAP50-95: mean Average Precision across IoU thresholds from 50% to 95% for 100 epochs.

| Class | Images | Instances | Precision | Recall | mAP50 | mAP50-95 |
|---|---|---|---|---|---|---|
| **All** | 1616 | 1685 | 0.975 | 0.965 | 0.989 | 0.811 |
| **basophil** | 201 | 201 | 1 | 0.994 | 0.995 | 0.833 |
| **eosinophil** | 167 | 168 | 1 | 0.996 | 0.995 | 0.912 |
| **erythroblast** | 223 | 241 | 0.975 | 0.979 | 0.99 | 0.736 |
| **Ig** | 193 | 208 | 0.916 | 0.942 | 0.974 | 0.833 |
| **lymphocyte** | 193 | 196 | 0.973 | 0.985 | 0.993 | 0.808 |
| **monocyte** | 179 | 179 | 0.975 | 0.978 | 0.993 | 0.81 |
| **neutrophil** | 213 | 218 | 0.957 | 0.928 | 0.987 | 0.853 |
| **platelet** | 247 | 274 | 1 | 0.916 | 0.989 | 0.705 |

Based on performance simulations for the YOLOv10 and YOLOv10n models for detecting blood cell types, the results show that the YOLOv10n model tends to perform better at higher training times. Comparing the different models at different epochs, the YOLOv10 model shows lower accuracy, Recall, and mAP for all IoU thresholds at ten epochs, indicating poor training and inefficient performance compared to YOLOv10n. As the YOLOv10n model is trained for 50 and 100 epochs, it significantly improves accuracy, Recall and mAP. At 100 epochs, the YOLOv10n model achieves the highest performance, accuracy, and recall, and it improves mAP50 and mAP50-95 values, indicating its ability to detect and classify red cells accurately. It demonstrates that an increased number of epochs and efficient YOLOv10n architecture give better results for detecting blood cells.

### 4.3 Comparative Analysis of different epoch configurations on MobileNetv2, ShuffleNetv2 and DarkNet for Blood Cell Classification

#### a. Comparative Analysis of MobileNetv2 for Blood Cell Classification for 10, 50 and 100 epochs

The model was trained first for ten epochs, at which point it reached an accuracy of 86.649%. The precision was 88.9685%, with a recall of 86.649%. This resulted in an F1 score of 86.6834% in total. This summarizes an excellent first learning phase with some space for further improvement of the model's predictive abilities. It showed remarkable improvement when trained for 50 epochs. The accuracy went up to about 94.3413%, with precision and recall going up to 95.3083% and 94.3413%, respectively. A far better F1 score was obtained at 94.4608%. This increase thus proves that the extended training drastically improves the model's ability to classify blood cells. Further training for 100 epochs brought marginal gains to 95.9328%. Precision and Recall were 96.1645% and 95.9329%, respectively; the F1 score was 95.8945%. Most of those metrics suggest a slight improvement over the 50-epoch train, although not as high, which again proves that the model converges and produces the optimal results during training around 50 epochs.

**Table 8**: Performance Metrics for Blood Cell Classification using MobileNetv2

| **Epochs** | **Accuracy** | **Precision** | **Recall** | **F1 Score** |
|---|---|---|---|---|
| **10 epochs** | 86.649 | 88.9685 | 86.649 | 86.6834 |
| **50 epochs** | 94.3413 | 95.3083 | 94.3413 | 94.4608 |
| **100 epochs** | 95.9328 | 96.1645 | 95.9329 | 95.8945 |

Table 8 and Figure 15 show that training MobileNetv2 for blood cell classification improves performance metrics up to 100 epochs, but most significantly up to 50. Beyond this point, gains are less substantive, indicating a possible flatlining of the model's learning curve. High precision and recall values at 50 and 100 epochs showed the robustness and reliability of this model for blood cell classification tasks.

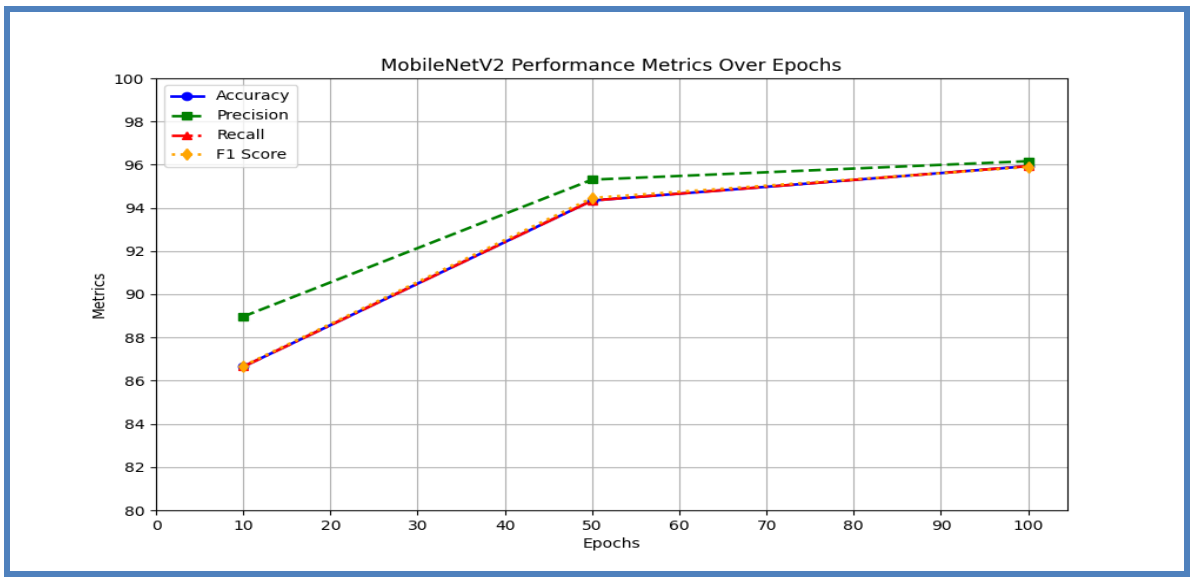

**Figure 15:** The MobileNetv2 Performance Metrics over 10, 50 and 100 Epochs.

b. **Comparative Analysis of ShuffleNetv2 for Blood Cell Classification for 10, 50 and 100 epochs**

After training for an initial ten epochs, the accuracy reached 92.22% with Precision and Recall of 93.84% and 92.23%, respectively, while the F1 score was 92.28%. These results show that ShuffleNetV2 is strong enough to learn from the data in the early stages of training itself and can act as a good initialization for further training. When the model was trained for 50 epochs, performance metrics showed reasonable improvements in accuracy at 94.96 per cent, precision at 95.35 per cent, and recall at 94.96 per cent. The F1 score improved to 94.99%. These improvements support the idea that, with extended training, the model becomes significantly better at correctly classifying blood cells. Further extension of the training to 100 epochs produced brilliant performance metrics: accuracy of 97.35%, Precision and Recall of 97.42% and 97.35%, respectively, and an F1 score of 97.36%. These results show that ShuffleNetV2 will keep fine-tuning its classification capability with increased training and hit high accuracy with reliability.

**Table 9:** Performance Metrics for Blood Cell Classification using ShuffleNetv2

| Epochs | Accuracy | Precision | Recall | F1 Score |
|---|---|---|---|---|
| 10 epochs | 92.22 | 93.84 | 92.23 | 92.28 |
| 50 epochs | 94.96 | 95.35 | 94.96 | 94.99 |
| 100 epochs | 97.35 | 97.42 | 97.35 | 97.36 |

In the result shown in Table 9 & Figure 16, ShuffleNetV2 is very effective for blood cell classification. Substantial improvements in the performance metrics are observed up to 100 epochs. An accuracy of 97.35% with a corresponding F1 score of 97.36% was attained at 100 epochs. Hence, ShuffleNetV2 can become a powerful instrument for medical image classification tasks since it can provide reliable and accurate results. The high precision proves this and recalls values across the different training epochs, which further reiterates that ShuffleNetV2 is robust for the blood cell image classification task.

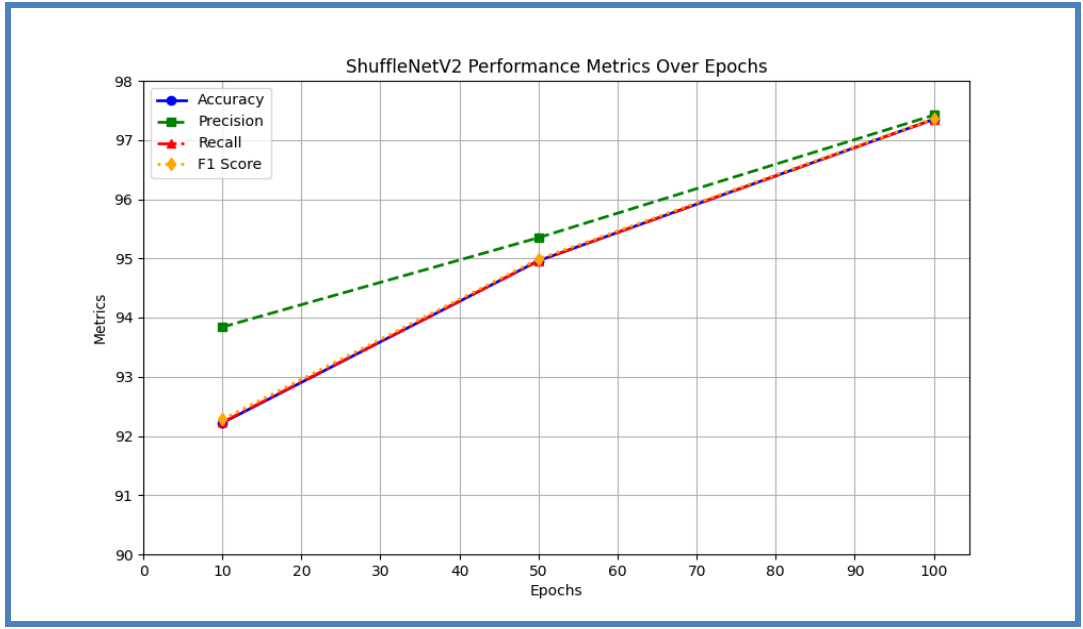

**Figure 16:** The ShuffleNetV2 Performance Metrics over 10, 50 and 100 Epochs.

c. **Comparative Analysis of DarkNet for Blood Cell Classification for 10, 50 and 100 epochs**

Training for the first ten epochs, DarkNet gave an accuracy of 79.58%. The precision is 81.92%, the Recall is 79.58%, and the F1 score is 78.17%. These early scores reflect that the model begins with a class of rather run-of-the-mill calculations, suggesting that more training is necessary to improve performance drastically. Training for 50 epochs has considerably increased performance. The accuracy increased to 93.8%, while the Precision and Recall went up to 94.03% and 93.8%, respectively. The F1 score also shot up to 93.65%. This dramatic improvement only reiterates that the larger the epochs, the better Darknet is at grasping the features critical to blood cell classification. Going further to 100 epochs in training, Darknet hit an accuracy level of 96.49%. The precision was 96.8%, Recall was 96.49%, and the F1 score was 96.47%. It suggests that DarkNet will perform better when trained for longer; the model achieves higher performance metrics, signifying the strength and practical ability to classify blood cells accurately.

**Table 10:** Performance Metrics for Blood Cell Classification using DarkNet

| Epochs | Accuracy | Precision | Recall | F1 Score |
|---|---|---|---|---|
| **10 epochs** | 79.58 | 81.92 | 79.58 | 78.17 |
| **50 epochs** | 93.8 | 94.03 | 93.8 | 93.65 |
| **100 epochs** | 96.49 | 96.8 | 96.49 | 96.47 |

The results in Table 10 and Figure 17 indicate that DarkNet is a robust model for categorizing blood cells, showing an improvement in performance metrics with an increase in training epoch. From a decent beginning accuracy of 79.58% at ten epochs, the model grew to 93.8% at 50 and 96.49% at 100 epochs. The results generally indicate that, given a more extended training period, DarkNet can fine-tune the classification ability. Thus, precision, Recall, and F1 scores were high at 100 epochs, proving the efficiency and reliability of the model in medical image classification tasks, making this tool invaluable for blood cell classification.

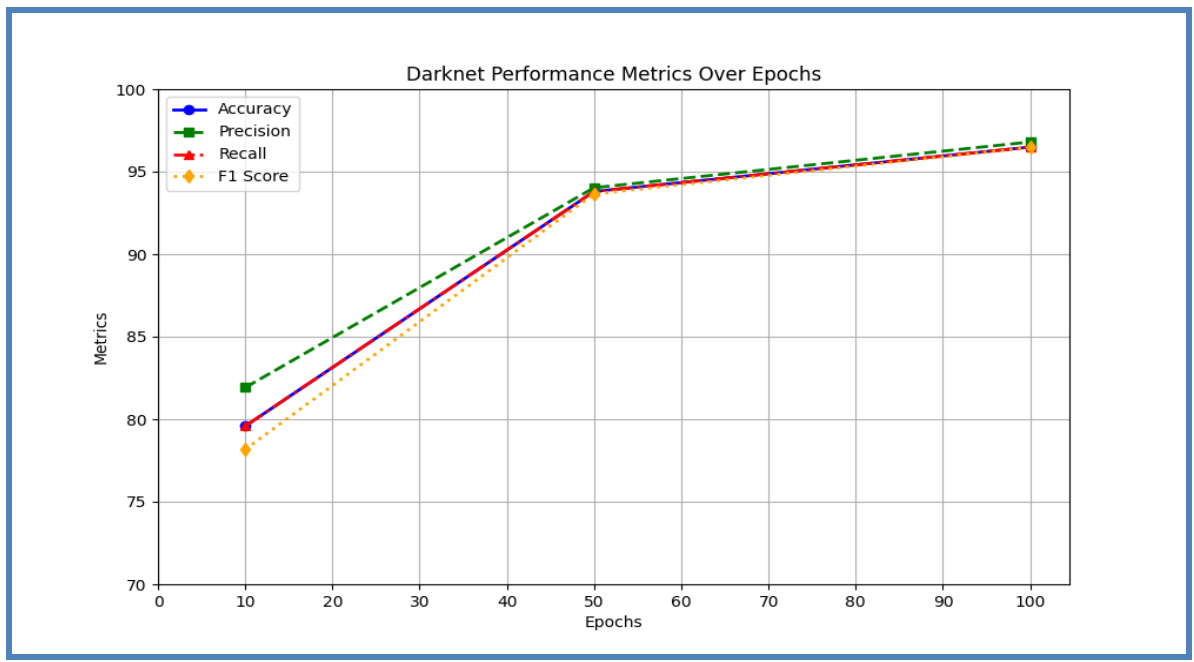

**Figure 17:** The DarkNet Performance Metrics over 10, 50 and 100 Epochs.

Based on blood cell classification performance metrics using three models—MobileNetv2, ShuffleNetv2, and DarkNet—MobileNetv2 and ShuffleNetv2 outperform DarkNet at higher epochs (50 and 100.) However, MobileNetv2 outperforms all metrics consistently, especially 100 epochs, with the highest accuracy (97.35%)., obtain Precision (97.42%), recall (97.35%), and F1 score (97.36%). This indicates that MobileNetv2 provides the best consistent classification performance, making it the best model for red blood cell classification, especially regarding accuracy. Regarding efficiency, ShuffleNetV2 also exhibits strong performance, especially in 100 epochs, but lags slightly behind MobileNetv2. DarkNet, while improving significantly with epochs, still lags behind the other models in terms of accuracy and Recall.

### 4.4 Discussions

There are several limitations for this study such as diversity and quality of datasets, computational requirements and the generalizability of the model to a variety of clinical scenarios. Although the dataset was thoroughly curated, briefed, and labeled, it might still be missing uncommon types of blood cells or pathological variations observed in clinical practice that might impact the model's ability to generalize to previously unseen clinical data. Image quality also is a key factor in detection and classification accuracy.

Noise, staining artifacts, low resolution inputs, etc. can all significantly hinder model performance. Future work could also utilize advanced image preprocessing methods like adaptive histogram equalization, noise reduction filters, and GAN-based image enhancement to make the model more robust against poor quality inputs. While the YOLOv10 model yielded promising results against high-resolution images (640×640 pixels), real-world applications may encounter reduced quality images, resulting from diverse sample preparation and microscope set ups. A better design in this regard would be to branch out in self-supervised learning or domain adaptation techniques to make the model easily adaptable to these changes. Moreover, the computational requirements of YOLOv10 may restrict its use in low-resource scenarios.

In addition, YOLOv10 may be resource intensive, especially when it comes to training and deployment of real-time applications. Alternative models such as MobileNetV2 and ShuffleNetV2 yield competitive performance, but the performance of YOLOv10 will undoubtedly outperform them in complex cases due to lower accuracy. Finally, since these models are based on the pre-defined architectures, it may well not adapt to clinical workflow or special needs of diagnosis based on some specific tunings.

## 5. Conclusions & Future Work

This paper focuses on the effectiveness of blood cell detection and classification by applying convolutional neural networks and YOLOv10. Our study indicates that the YOLOv10n model achieves the highest Accuracy, Precision, Recall, and mAP values, particularly at 100 epochs, demonstrating its effectiveness in accurate detection. Additionally, the F1-score was considered, as it provides a balanced evaluation by integrating Precision and Recall, making it a more reliable metric for imbalanced datasets. While the dataset used in this study exhibits a relatively balanced distribution, with class sizes ranging from 160 to 250 samples, the F1-score was utilized due to its robustness in evaluating model performance beyond accuracy alone. Accuracy can be misleading; for instance, in a dataset where 95% of samples are negative, a model predicting all negative cases would achieve 95% accuracy while failing to detect the positive cases. The F1-score ensures an optimal trade-off between Precision (minimizing false positives) and Recall (reducing false negatives). This balance is particularly crucial in medical applications, where high recall mitigates the risk of missed diagnoses, and high precision reduces unnecessary medical interventions, making the F1-score a more reliable performance metric in this domain. For red blood cell classification, the MobileNetv2 model provides consistently good performance across all metrics, especially at 100 epochs, where it outshines both ShuffleNetv2 and DarkNet regarding high accuracy, Precision, and Recall. MobileNetv2 appears as the best model for blood cell classification. ShuffleNetv2 also performs well but still lags slightly behind MobileNetv2 in accuracy. DarkNet, while showing improvement with increasing epochs, performs relatively lower than the other models in terms of overall classification accuracy. Results in classification showed outstanding performance, stating advantages brought by the deep learning process in medical image analysis. Advanced techniques improved the model's capacity to identify complex patterns and reduce over-fitting.

Currently, this study is limited to only one dataset of blood cell images. This limited the degree of our analysis and, therefore, the generalizability of our findings. Further work will enable a deeper look into model performance when more extensive and diverse datasets become available. These models can also be utilized in IoT research to integrate with microscopes or hemocytometers, enabling real-time blood cell analysis and automated diagnostics. Additionally, if a suitable dataset becomes available in the future, aiming to incorporate age-group-based analysis to examine variations in blood cell morphology across different demographic groups,further enhances the model's applicability.

**Acknowledgements**

This research was financially supported by Princess Nourah bint Abdulrahman University Researchers Supporting Project number (PNURSP2025R235), Princess Nourah bint Abdulrahman University, Riyadh, Saudi Arabia.

**Funding**

This research was financially supported by Princess Nourah bint Abdulrahman University Researchers Supporting Project number (PNURSP2025R235), Princess Nourah bint Abdulrahman University, Riyadh, Saudi Arabia.

**Data availability**

All the datasets utilized for experimentation in this study are publicly available at: **https://www.kaggle.com/datasets/siddxt/blood-cells-annotated-yolo-dataset**